\documentclass[letterpaper, 10 pt, conference]{ieeeconf}  % Comment this line out
\IEEEoverridecommandlockouts                              % This command is only
\usepackage[utf8]{inputenc}
\usepackage[T1]{fontenc}
\usepackage{lipsum,booktabs}
\usepackage{amsmath}
\usepackage{graphicx}
\usepackage{algorithm,algorithmic}
\usepackage{multirow}
\usepackage{subfig}
\usepackage{fancyhdr}
\usepackage{xspace}

\title{\LARGE \bf
NeuroMorphix: A Novel Brain MRI Asymmetry-specific Feature Construction Approach For Seizure Recurrence Prediction
}

\author{Soumen Ghosh, Viktor Vegh, Shahrzad Moinian, Hamed Moradi, Alice-Ann Sullivan, John Phamnguyen, \\David Reutens % <-this % stops a space
\thanks{S. Ghosh, V. Vegh, S. Moinian, H. Moradi, J. Phamnguyen and D. Reutens are with the Centre for Advanced Imaging, The University of Queensland, Australia (e-mail: soumen.ghosh@uq.edu.au, viktor.vegh@cai.uq.edu.au, d.reutens@uq.edu.au).}
\thanks{S. Ghosh, V. Vegh, S. Moinian, H. Moradi and D. Reutens are with the ARC-funded Training Centre for Innovation in Biomedical Imaging Technology, The University of Queensland, Australia.}
\thanks{AA. Sullivan, J. Phamnguyen and D. Reutens are with the  Royal Brisbane and Women's Hospital, Brisbane, Australia (e-mail: alice-ann.sullivan@health.qld.gov.au).}}%

\begin{document}
\maketitle
\thispagestyle{plain}
\pagestyle{plain}

\let\thefootnote\relax\footnotetext{\textbf{This work has been submitted to the IEEE for possible publication. Copyright may be transferred without notice, after which this version may no longer be accessible.}}

%%%%%%%%%%%%%%%%%%%%%%%%%%%%%%%%%%%%%%%%%%%%%%%%%%%%%%%%%%%%%%%%%%%%%%%%%%%%%%%%
\section*{Abstract}
Seizure recurrence is an important concern after an initial unprovoked seizure; without drug treatment, it occurs within 2 years in 40-50\% of cases. The decision to treat currently relies on predictors of seizure recurrence risk that are inaccurate, resulting in unnecessary, possibly harmful, treatment in some patients and potentially preventable seizures in others. Because of the link between brain lesions and seizure recurrence, we developed a recurrence prediction tool using machine learning and clinical 3T brain MRI. We developed NeuroMorphix, a feature construction approach based on MRI brain anatomy. Each of seven NeuroMorphix features measures the absolute or relative difference between corresponding regions in each cerebral hemisphere. FreeSurfer was used to segment brain regions and to generate values for morphometric parameters (8 for each cortical region and 5 for each subcortical region). The parameters were then mapped to whole brain NeuroMorphix features, yielding a total of 91 features per subject. Features were generated for a first seizure patient cohort (n = 169) categorised into seizure recurrence and non-recurrence subgroups. State-of-the-art classification algorithms were trained and tested using NeuroMorphix features to predict seizure recurrence. Classification models using the top 5 features, ranked by sequential forward selection, demonstrated excellent performance in predicting seizure recurrence, with area under the ROC curve of 88-93\%, accuracy of 83-89\%, and F1 score of 83-90\%. Highly ranked features aligned with structural alterations known to be associated with epilepsy. This study highlights the potential for targeted, data-driven approaches to aid clinical decision-making in brain disorders.

\begin{keywords}
Brain MRI, brain-specific feature, seizure recurrence prediction, feature construction approach, machine learning
\end{keywords}

\section{Introduction}
Accurate prediction of seizure recurrence is important after a first unprovoked seizure because of the potential for injury or death from recurrent seizures. Currently, antiseizure medications are started in individuals predicted to be at elevated risk of seizure recurrence. However, currently used clinical, electroencephalography (EEG) and magnetic resonance imaging (MRI) predictors are unreliable, being associated with a recurrence rate of 60\% \cite{lawn2015first} compared to 25\% if these risk factors are absent \cite{berg1991risk} \cite{krumholz2015evidence}. As a result, a significant fraction of individuals receiving treatment are unnecessarily exposed to the side effects of medication and conversely, a significant fraction of those who remain untreated are exposed to the risk of a further seizure.

The search for better predictive models utilising machine learning is active and ongoing. A predictive model developed using clinical data and EEG findings from the most extensive study to date on individuals experiencing a single seizure, the Multicentre trial for Early Epilepsy and Single Seizures (MESS) only achieved a discriminatory power slightly higher than a random guess (C-statistic: 0.59) \cite{kim2006prediction}. The Area under the Received Operating Characteristic curve (AUC) for other seizure recurrence prediction models using clinical information and EEG has ranged between 0.63 \cite{lemoine2023machine} and 0.86 \cite{van2018prediction}. Although a recent Large Language Model utilising unstructured electronic medical records for seizure recurrence prediction achieved an AUC of 90\% \cite{beaulieu2023predicting}, training inputs included information on medication and prognosis and could also have been biased by the expertise of clinicians. 

Abnormal brain imaging is a known risk factor for seizure recurrence \cite{berg1991risk, gavvala2016new, wiebe2008evidence, lapalme2022neuroimaging}. The hazard ratio for seizure recurrence in individuals with abnormal versus normal neuroimaging is 2.44, higher than that for individuals with epileptiform abnormalities on the EEG versus those with a normal EEG (2.16)  \cite{krumholz2015evidence}. Detection of an epileptogenic lesion of the cerebral cortex on MRI is associated with seizure recurrence in 67\% of cases \cite{ho2013neuroimaging}. While lesions such as hippocampal sclerosis \cite{semah1998underlying}, malformations of cortical development \cite{chen2019histological, guerrini2014malformations} and tumors are often amenable to diagnosis by visual inspection, a range of epileptogenic pathologies are difficult to detect in this manner \cite{zhao2019reduced, voets2011increased, lin2007reduced, ronan2007cerebral, sisodiya1995widespread}. Machine learning algorithms have shown promising results in medical image segmentation \cite{seo2020machine}, classification \cite{koitka2016traditional}, and diagnosis \cite{myszczynska2020applications} with recent studies \cite{guo2015automated, ong2022detection} showcasing effectiveness in detecting subtle lesions on brain MRI scans. This raises the possibility of increasing the accuracy of seizure recurrence prediction by developing methods that are sensitive to subtle or currently invisible abnormalities. Since developmental and acquired disorders associated with epilepsy are known to cause deviations from the typical pattern of asymmetry between the cerebral hemispheres \cite{shah2019structural, cook1992hippocampal}, we developed NeuroMorphix, an approach involving the construction of brain asymmetry-specific features using 3D brain MRI data and applied it to predict seizure recurrence using state-of-the-art classification algorithms. NeuroMorphix features collectively quantify differences between cerebral hemispheres based on a range of cortical and subcortical region-specific measures. The classification algorithms learn patterns from the whole-brain features generated from clinical brain MRI to classify patients into those with and without recurrent seizures after a first unprovoked seizure.

\section{Methods} \label{mri_model}
\begin{figure*}
    \centering
    \includegraphics[scale=0.65]{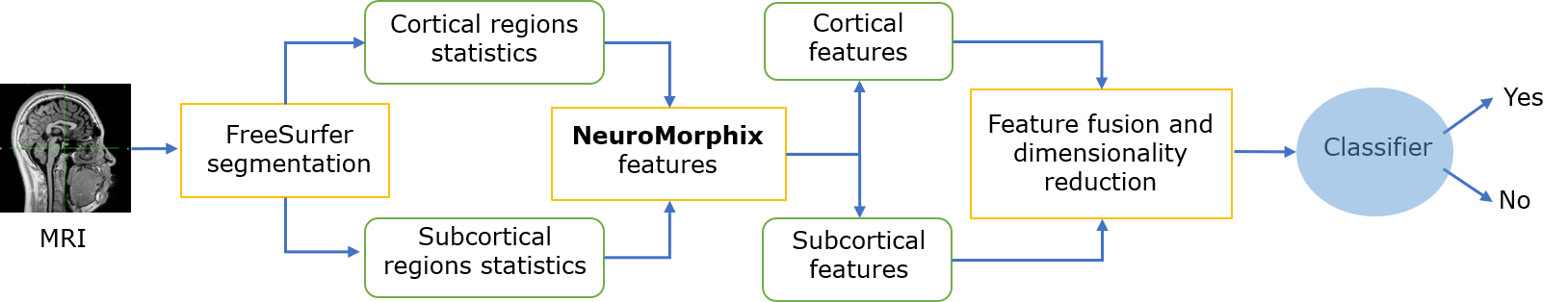}
    \caption{Proposed framework for seizure recurrence prediction using proposed NeuroMorphix features.}
    \label{fig:firstsz_framework}
\end{figure*}

We used FreeSurfer (Version 7.1.1; \cite{fischl2012freesurfer}) to generate a set of standard region-specific parameters from which NeuroMorphix features were created. These features were used to perform the classification task. Figure \ref{fig:firstsz_framework} summarises the steps in this process.

\subsection{Region-specific parameters}
These were calculated for cortical and subcortical regions.

\begin{table}[]
\centering
\begin{tabular}{|l|l|l|}
\hline
\multicolumn{1}{|c|}{\textbf{Sl No}} & \multicolumn{1}{c|}{\textbf{FreeSurfer Parameter Name}} & \multicolumn{1}{c|}{\textbf{Label}} \\ \hline
1                                    & Surface Area (in $mm^2$)                                        & SurfArea                              \\ \hline
2                                    & Gray Matter Volume (in $mm^3$)                                  & GrayVol                               \\ \hline
3                                    & Average Thickness (in $mm$)                                   & ThickAvg                              \\ \hline
4                                    & Thickness StdDev (in $mm$)                                   & ThickStd                              \\ \hline
5                                    & Integrated Rectified Mean   Curvature               & MeanCurv                              \\ \hline
6                                    & Integrated Rectified Gaussian   Curvature           & GausCurv                              \\ \hline
7                                    & Folding Index                                       & FoldInd                               \\ \hline
8                                    & Intrinsic Curvature Index                           & CurvInd                               \\ \hline
\end{tabular}
\caption{Cortical parameters computed by FreeSurfer.}
\label{tab:cortical_stats}
\end{table}

\begin{table}[]
\centering
\begin{tabular}{|l|l|l|}
\hline
\multicolumn{1}{|c|}{\textbf{Sl No}} & \multicolumn{1}{c|}{\textbf{FreeSurfer Parameter Name}} & \multicolumn{1}{c|}{\textbf{Label}} \\ \hline
1                                    & Volume (in $mm^3$)                                             & Volume                                \\ \hline
2                                    & Number of Voxels                                       & NVoxels                               \\ \hline
3                                    & Intensity normMean                                     & normMean                              \\ \hline
4                                    & Intensity normStdDev                                   & normStdDev                            \\ \hline
5                                    & Intensity normMax                                      & normMax                               \\ \hline
\end{tabular}
\caption{Subcortical parameters computed by FreeSurfer.}
\label{tab:subcortical_stats}
\end{table}

\subsubsection{Cortical parameters}
We considered curvature index, folding index, Gaussian curvature, mean curvature, thickness, thickness standard deviation, and volume of cortical regions, as listed in Table \ref{tab:cortical_stats}. The cortical parameters were stored as $X_{R}$ and $X_{L}$, matrices of dimension $C \times P$, where $C$ is the number of cortical regions, and $P$ is the number of region-specific parameters generated for the right ($R$) and left ($L$) cerebral hemispheres.

\subsubsection{Subcortical parameters}
These parameters included maximum intensity, mean intensity, standard deviation of intensity, and volume, see Table \ref{tab:subcortical_stats}, which were also stored in the $X_{R}$ and $X_{L}$ format.

\subsection{NeuroMorphix features}
Seven asymmetry features, $f_{1,i}, f_{2,i}, ..., f_{7,i}$, were calculated, reflecting cosine similarity, outlier count, and hemispheric ratio for each region-specific parameter, $i$ = $1,..,M$.

\subsubsection{Features based on cosine similarity}
We defined $f_{1,i}$ as the cosine similarity between corresponding regions in each cerebral hemisphere:
\begin{equation}
    f_{1,i} = \frac{X_{L,i} \cdot X_{R,i}}{||X_{L,i}||||X_{R,i}||},
    \label{eq:f1}
\end{equation}
where $i$ denotes a specific column of the matrix (i.e., region-specific parameter).

Feature $f_{2,i}$ was based on the normalised absolute deviation from the mean of all the brain regions considered:

\begin{equation}
    Z_{L,i} = \frac{|\textbf{u}\overline{X}_{L,i} - X_{L,i}|}{\overline{X}_{L,i}},
\end{equation}

\begin{equation}
    Z_{R,i} = \frac{|\textbf{u}\overline{X}_{R,i} - X_{R,i}|}{\overline{X}_{R,i}},
\end{equation}

where $Z_{L,i}$ and $Z_{R,i}$ are vectors corresponding to the left and right hemisphere regions, $\textbf{u}$ is a vector of all ones and, for example, $\overline{X}_{L,i}$ is the mean of $X_{L,i}$. Feature $f_{2,i}$ measured the similarity between $Z_{L,i}$ and $Z_{R,i}$ according to:

\begin{equation}
    f_{2,i} = 
    Z_{L,i} \cdot Z_{R,i}.
    \label{eq:f2}
\end{equation}

\subsubsection{Feature based on outlier count}
Features $f_{3,i}$ and $f_{4,i}$  were the ratios of the number of regions outside a specified cutoff defined as $\overline{X} \pm \varepsilon$, where $\varepsilon$ was chosen to suit the problem. Here, following common practice statistical outlier detection, we considered $\varepsilon$ to be the standard deviation of $X_{L,i}$ or $X_{R,i}$. Feature $f_{3,i}$ was calculated from the ratio between the number of regions above the cutoff in each hemisphere:

\begin{equation}
    f_{3,i} = \text{min}\left(\frac{y_{L,i}} {y_{R,i}}, \frac{y_{R,i}}{y_{L,i}}\right),
    \label{eq:f3}
\end{equation}
where $y_{L,i}$ = $\text{count}(X_{L,i,j}|_{j=1,..,N} > (\overline{X}_{L,i} + \varepsilon))$, $y_{R,i}$ = $\text{count}(X_{R,i,j}|_{j=1,..,N} > (\overline{X}_{R,i} + \varepsilon))$, $N$ is the number of brain regions considered, $y_{L,i}$ and $y_{R,i}$ take integer values, and $f_{3,i}$ is a scalar between 0 and 1.

Similarly, $f_{4,i}$ related to the case when the cutoff was set at a threshold below the mean: 
\begin{equation}
    f_{4,i} = \text{min}\left(\frac{y_{L,i}} {y_{R,i}}, \frac{y_{R,i}}{y_{L,i}}\right).
    \label{eq:f4}
\end{equation}
where $y_{L,i} = \text{count}(X_{L,i,j}|_{j=1,..,N} < (\overline{X}_{L,i} - \varepsilon))$ and $y_{R,i} = \text{count}(X_{R,i,j}|_{j=1,..,N} < (\overline{X}_{R,i} - \varepsilon))$.

\subsubsection{Ratio features}
These features involve the ratio of region-specific parameters. We first computed, using entry-wise division, a vector of ratios:

\begin{equation}
    \textbf{r}_i = \left[\text{min} \left( X_{L,i,j} \div X_{R,i,j}|_{j=1,..,N} \right)\right],
\end{equation}
where $\textbf{r}_i$ is of length $N$, and entries in $\textbf{r}_i$ lie  between 0 and 1. The features were computed as:
\begin{equation}
    f_{5,i} = \mu(\textbf{r}_i),
    \label{eq:f5}
\end{equation}
\begin{equation}
    f_{6,i} = \sigma(\textbf{r}_i),
    \label{eq:f6}
\end{equation}
\begin{equation}
    f_{7,i} = \text{min}(\textbf{r}_i).
    \label{eq:f7}
\end{equation}
where $\mu$ is the arithmetic mean operator, and $\sigma$ is the standard deviation.

\section{Evaluation Dataset}
The study was approved by the institutional Human Research Ethics Committee (LNR/2019/QRBW/55712). We collected retrospective brain MRI scans for first seizure patients seen from March 2016 to October 2020 at the Epilepsy Clinic of the Royal Brisbane and Women’s Hospital, Brisbane, Australia. A total of 169 brain MRI datasets were collected and categorised into those from patients with ($n = 145$) and without ($n = 24$) recurrent seizures. The cohort comprised 102 males with an average age at the time of first seizure of 37.7 ($\pm 16.3$) years.
Scans were collected using a 3T MRI scanner (Magnetom Vida, Siemens HealthCare). Our analysis utilised $\text{T}_1$-weighted images (TR = 2.3$s$, TE = 2.31$ms$, matrix = $256\times256$, $\alpha = 8^{\circ}$) with a resolution of either $0.9 mm^3$ (132 scans) or $0.9\times 0.45\times 0.45 mm^3$ (37 scans). 
The datasets were resampled using trilinear interpolation to a common resolution of $1.0 mm^3$ and matrix size of $256 \times 256 \times 192$ using the Medical Image Processing, Analysis and Visualisation (MIPAV, Version 10.0.0) software package.

\section{Experiment details}
The FreeSurfer (Version 7.1.1) software package was used for brain segmentation and region-specific parameter generation, and machine learning approaches were implemented using Python (Version 3.9.16). Multi-dimensional matrix operations were performed using the NumPy (Version 1.23.5) and pandas (Version 1.5.2) libraries. Tools necessary to perform sequential forward feature selection and sequential backward elimination were imported from scikit (Version 1.2.1). Algorithms were implemented and run on Linux-centos7 OS-based computing systems.

\subsection{FreeSurfer parameters}
Each MRI, in native space, was segmented into cortical and subcortical regions according to the Desikan-Killiany atlas using the recon all command in FreeSurfer. Cortical parameters were obtained from the $lh.aparc.stats$ and $rh.aparc.stats$ files for each hemisphere and subcortical parameters were obtained from the $aseg.stats$ file for both hemispheres.

Eight parameters (see Table \ref{tab:cortical_stats}) for 34 cortical regions were available for each hemisphere in all participants. Using the $aparcstats2table$ command, the $aparc$ parameters of all subjects were organised in eight $csv$ files, each representing one FreeSurfer parameter for each hemisphere. The dimension of each $csv$ file was $169\times34$, in keeping with 34 cortical regions in 169 participants. The 5 parameters (see Table \ref{tab:subcortical_stats}) for 14 subcortical regions in each hemisphere were similarly organised in 5 $csv$ files of dimension $169 \times 14$.

\subsection{Feature construction}
Each pair of corresponding left and right hemisphere $csv$ files was used to generate NeuroMorphix features (i.e., $f_1$ to $f_7$), yielding 56 cortical features (seven NeuroMorphix features for each of the eight FreeSurfer parameters) and 35 subcortical features (seven NeuroMorphix features for each of the five FreeSurfer parameters). These features were separately concatenated for cortical and subcortical regions, yielding a cortical NeuroMorphix feature set of dimension $169 \times 56$ and a subcortical NeuroMorphix feature set of dimension $169 \times 35$.

\subsection{Class imbalance}
To upsample the minority class prior to classification, we employed the Synthetic Minority Over-sampling Technique (SMOTE) \cite{chawla2002smote}, which adds synthetic samples to the minority class by interpolating between existing instances and their K-nearest neighbours. We used the SMOTE algorithm, as implemented in Python ($imblearn$ library Version 0.10.1). SMOTE was chosen because it generalises well and is less prone to overfitting than upsampling methods such as random sampling \cite{wongvorachan2023comparison}.

\subsection{Feature selection}
The total number of features (91) is large relative to the number of available datasets. We evaluated sequential forward selection and sequential backward elimination approaches for dimensionality reduction to prevent the robustness of the classification model from being compromised \cite{ferri1994comparative}. Feature selection was performed using the $SequentialFeatureSelector$ function in the Sklearn Python library (Version 1.2.1). The feature subset that achieved the highest classification accuracy for all classifiers was selected for use with the classification models.

\subsection{Classification models}
Being agnostic to the choice of classification model,  we evaluated the performance of a range of state-of-the-art machine learning approaches for classification: K-Nearest Neighbours (KNN; \cite{fix1952discriminatory}), Decision Tree (DT; \cite{loh2011classification}), Random Forest (RF; \cite{ho1995random}), Gradient Boosting (GB; \cite{friedman2001greedy}), Extreme Gradient   Boosting (XGB; \cite{chen2016xgboost}), and Light Gradient Boosting Machine (LGBM; \cite{ke2017lightgbm}) classifiers. 

The upsampled dataset was randomly split into training and test sets in a ratio of 70:30. A one-dimensional vector of length 290 (145 samples in each class) was provided as the class label. We used the default classifier function from the sk-learn machine learning library. The specifications of the classifiers used for training and testing were K $ = 3$ for KNN, depth $ = 4$ for RF, learning rate of 1.0 for GB and $gbdt$ boosting type for LGBM classifiers. In each case, 5-fold cross-validation was performed.

\subsection{Performance evaluation}
The following metrics were computed from the number of true positives (TP), true negatives (TN), false positives (FP), and false negatives (FN) \cite{sokolova2009systematic}:
\begin{equation}
    Accuracy = (TP + TN) / (TP + TN + FP + FN),
\end{equation}
\begin{equation}
    Specificity = TN/(FP+TN),
\end{equation}
\begin{equation}
    Sensitivity = TP / (TP + FN).
\end{equation}
\begin{equation}
    \text{F}_1 =  2 \times TP/(FP+2 \times TP+FN).
\end{equation}
The Area Under the Receiver Operator Characteristic (AUROC) curve was also used to assess the performance of each model.

\section{Results}
We report the results of five experiments. First, we ascertained the optimal number of features to be used with each classification algorithm. Second, we considered the case of using a fixed number of features and assessed the effects of the performance of each method. Third, we investigated the importance of individual features for each classification method. Fourth, 5-fold cross-validation was performed. Fifth, performance was evaluated in a test dataset to establish how well each model generalises.

\subsection{Optimal number of features}
\begin{figure*}
    \centering
    \subfloat[\centering Sequential forward selection]{{\includegraphics[scale=0.38]{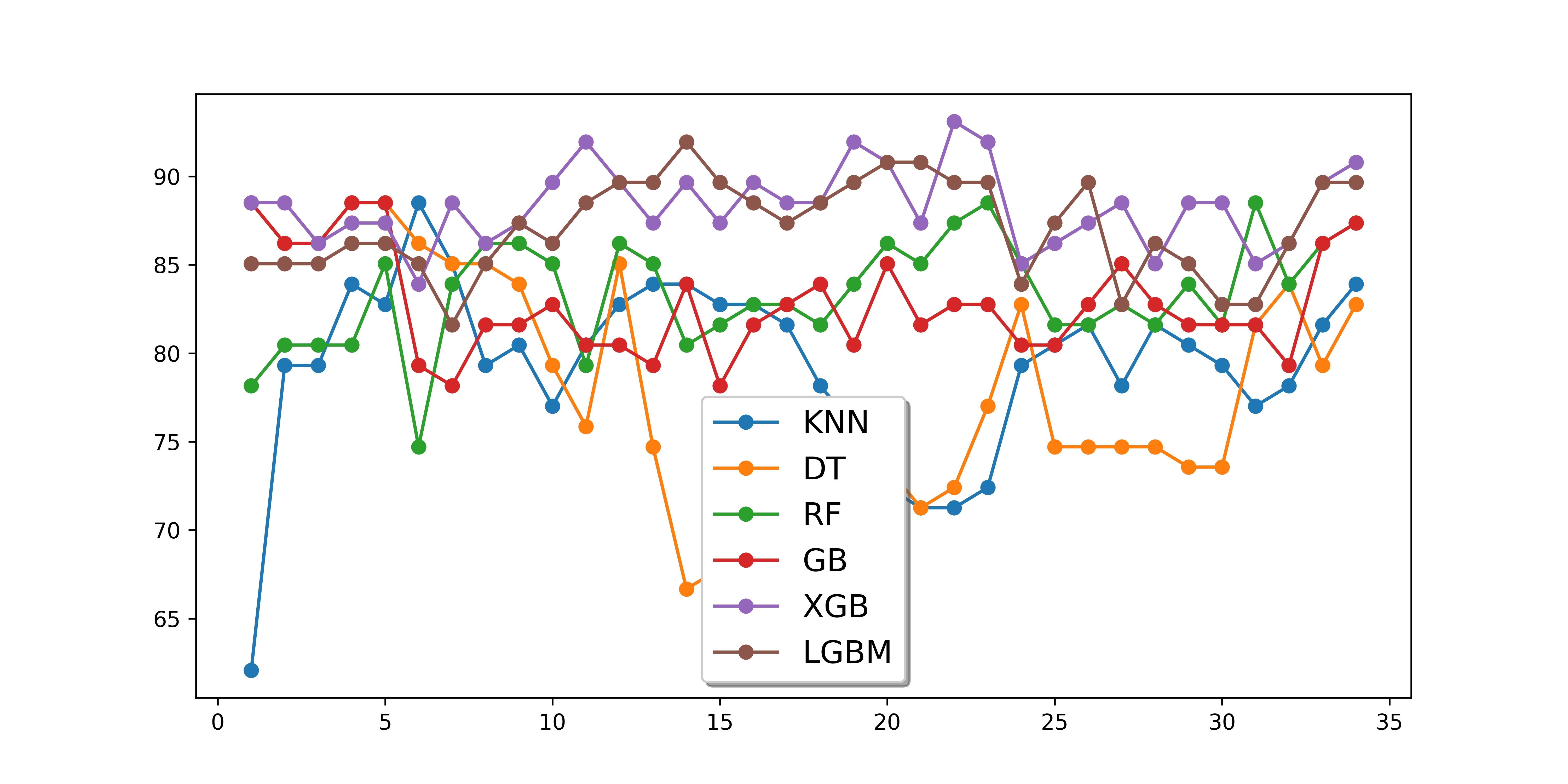}}}
    \subfloat[\centering Sequential backward elimination]{{\includegraphics[scale=0.38]{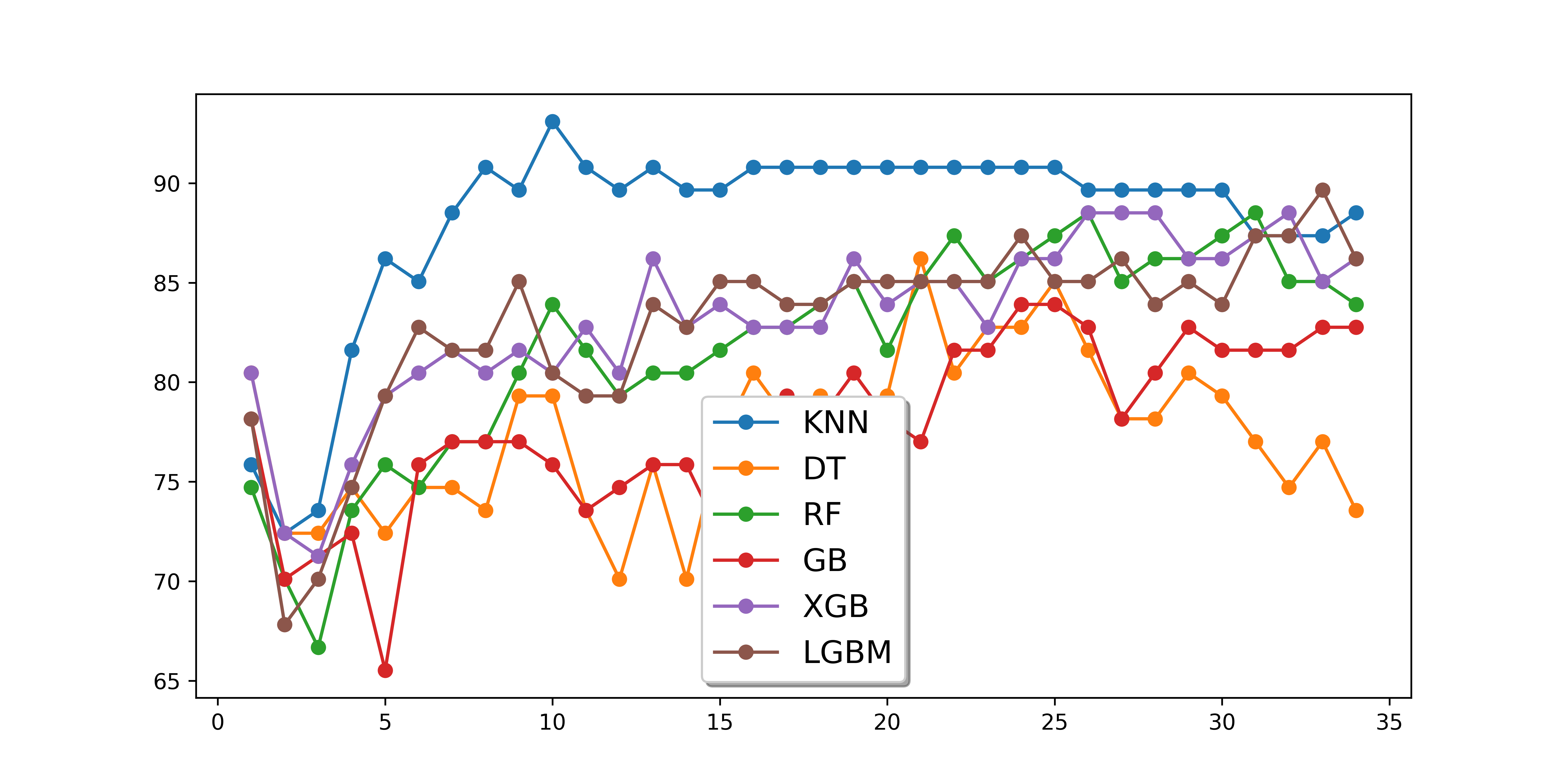}}}%
    \caption{Shown are feature selection results using the (a) sequential feature selection and (b) sequential backward elimination approaches. The horizontal axis represents the number of features used to achieve that level of accuracy (vertical axis). Colours correspond with the different classification algorithms evaluated.}
    \label{fig:sfs}
\end{figure*}

KNN (K $= 3$) was used for the sequential forward selection and backward elimination steps. Results are presented in Figure \ref{fig:sfs}, in which the horizontal and vertical axes respectively represent the number of features and the achieved classification accuracy. The purpose of this analysis was to identify the number of features above which classification accuracy does not significantly improve. For example, for XGB with the use of sequential forward selection, around 12 features were needed to obtain the best classification accuracy. However, around half as many features was required when KNN was used for classification. Of note, classifiers required a different number of NeuroMorphix features to achieve their highest classification accuracy. Generally, more features were required to achieve high classification accuracy when sequential backward elimination was used for feature selection. Other than KNN, classifiers also appeared to perform better when sequential forward selection was applied. While nearly 90\% classification accuracy was achieved using KNN with forward selection, classification accuracy exceeded 90\% for XGB and LGBM with forward selection. This was not the case when sequential backward elimination of features was used. We therefore used sequential forward selection in the final method.

Table \ref{tab:mri_model_multi_features} summarises the number of features selected using sequential forward selection and the corresponding performance metrics of each classification method. As few as two features can be used to achieve relatively good performance with DT and XGB, whereas KNN and LGBM require six and eight features, respectively, to achieve their best classification performance. RF and GB required 28 and 20 features, respectively, to perform well. One advantage of using a low number of features for classification is that interpretation of the features may be simpler than for a large feature set.

\begin{table*}[]
\centering
\begin{tabular}{|l|l|l|l|l|l|l|}
\hline
\multicolumn{1}{|c|}{\textbf{Algorithm}} & \textbf{Number of features} & \multicolumn{1}{c|}{\textbf{AUROC}} & \multicolumn{1}{c|}{\textbf{Accuracy}} & \multicolumn{1}{c|}{\textbf{Sensitivity}} & \multicolumn{1}{c|}{\textbf{Specificity}} & \multicolumn{1}{c|}{\textbf{F1}} \\ \hline
KNN                                       & 6                       & \textbf{94.27}                      & \textbf{88.51}                         & 90.70                                     & \textbf{86.36}                            & 88.36                        \\ \hline
DT                                        & 2                       & 86.58                               & 87.36                                  & \textbf{97.67}                            & 77.27                                     & \textbf{88.42}                        \\ \hline
RF                                        & 28                      & 90.17                               & 85.06                                  & 86.05                                     & 84.09                                     & 85.06                                 \\ \hline
GB                                        & 20                      & 92.92                               & 82.76                                  & 79.07                                     & 86.36                                     & 81.93                                 \\ \hline
XGB                                       & 2                       & 91.49                               & 85.06                                  & 88.37                                     & 81.82                                     & 85.39                                 \\ \hline
LGBM                                      & 8                       & 93.13                               & 81.61                                  & 76.74                                     & 86.36                                     & 80.49                                 \\ \hline
\end{tabular}
\caption{Classification performance of the classifier assessed using sequential forward selection of features. The highest number in each column is denoted in \textbf{bold} font.}
\label{tab:mri_model_multi_features}
\end{table*}

\subsection{Fixed number of features}
To assess the variability between methods when the dimension of the feature space was fixed, we considered the cases of five NeuroMorphix features identified by sequential forward selection and of 25 features identified by sequential backward elimination. The results are presented in Table \ref{tab:sfs_sbe}. High performance could be achieved using the forward selected features (around 88\% accuracy for DT, RF, GB and XGB and 100\% sensitivity for DT, GB and LGBM). Accuracy and sensitivity were lower with sequential backward elimination. Note that in Figure \ref{fig:sfs}a, the number of features required by XGB and LGBM to achieve high classification accuracy plateaus quickly, and additional features only marginally changed the performance metrics. For KNN, performance degraded progressively as the number of sequential forward selected features increased.

\begin{table*}[]
\centering
\scalebox{1}{
\begin{tabular}{|l|lllll|lllll|}
\hline
\multicolumn{1}{|c|}{\multirow{2}{*}{\textbf{Algorithm}}} & \multicolumn{5}{c|}{\textbf{Sequential forward selection: 5 features}}                                                                                                                                                & \multicolumn{5}{c|}{\textbf{Sequential backward elimination: 25 features}}                                                                                                                                              \\ \cline{2-11} 
\multicolumn{1}{|c|}{}                                     & \multicolumn{1}{c|}{\textbf{AUROC}} & \multicolumn{1}{c|}{\textbf{Accuracy}} & \multicolumn{1}{c|}{\textbf{Sensitivity}} & \multicolumn{1}{c|}{\textbf{Specificity}} & \multicolumn{1}{c|}{\textbf{F1}} & \multicolumn{1}{c|}{\textbf{AUROC}} & \multicolumn{1}{c|}{\textbf{Accuracy}} & \multicolumn{1}{c|}{\textbf{Sensitivity}} & \multicolumn{1}{c|}{\textbf{Specificity}} & \multicolumn{1}{c|}{\textbf{F1}} \\ \hline
KNN                                                        & \multicolumn{1}{l|}{89.01}          & \multicolumn{1}{l|}{82.76}             & \multicolumn{1}{l|}{88.37}                & \multicolumn{1}{l|}{77.27}                & 83.52                                 & \multicolumn{1}{l|}{89.32}          & \multicolumn{1}{l|}{83.91}             & \multicolumn{1}{l|}{72.09}                & \multicolumn{1}{l|}{\textbf{95.45}}       & 81.58                                 \\ \hline
DT                                                         & \multicolumn{1}{l|}{88.74}          & \multicolumn{1}{l|}{\textbf{88.51}}    & \multicolumn{1}{l|}{\textbf{100.00}}      & \multicolumn{1}{l|}{77.27}                & \textbf{89.58}                        & \multicolumn{1}{l|}{81.50}          & \multicolumn{1}{l|}{81.61}             & \multicolumn{1}{l|}{72.09}                & \multicolumn{1}{l|}{90.91}                & 79.49                                 \\ \hline
RF                                                         & \multicolumn{1}{l|}{88.66}          & \multicolumn{1}{l|}{\textbf{88.51}}    & \multicolumn{1}{l|}{90.70}                & \multicolumn{1}{l|}{\textbf{79.55}}       & 85.71                                 & \multicolumn{1}{l|}{92.76}          & \multicolumn{1}{l|}{81.61}             & \multicolumn{1}{l|}{81.40}                & \multicolumn{1}{l|}{81.82}                & 81.40                                 \\ \hline
GB                                                         & \multicolumn{1}{l|}{\textbf{92.55}} & \multicolumn{1}{l|}{\textbf{88.51}}    & \multicolumn{1}{l|}{\textbf{100.00}}      & \multicolumn{1}{l|}{77.27}                & \textbf{89.58}                        & \multicolumn{1}{l|}{86.31}          & \multicolumn{1}{l|}{85.06}             & \multicolumn{1}{l|}{79.07}                & \multicolumn{1}{l|}{90.91}                & 83.95                                 \\ \hline
XGB                                                        & \multicolumn{1}{l|}{89.90}          & \multicolumn{1}{l|}{87.36}             & \multicolumn{1}{l|}{95.35}                & \multicolumn{1}{l|}{\textbf{79.55}}                & 88.17                                 & \multicolumn{1}{l|}{\textbf{95.08}} & \multicolumn{1}{l|}{\textbf{89.66}}    & \multicolumn{1}{l|}{\textbf{83.72}}       & \multicolumn{1}{l|}{\textbf{95.45}}       & \textbf{88.89}                        \\ \hline
LGBM                                                       & \multicolumn{1}{l|}{87.55}          & \multicolumn{1}{l|}{86.21}             & \multicolumn{1}{l|}{\textbf{100.00}}               & \multicolumn{1}{l|}{72.73}                & 87.76                                 & \multicolumn{1}{l|}{92.81}          & \multicolumn{1}{l|}{85.06}             & \multicolumn{1}{l|}{81.40}                & \multicolumn{1}{l|}{88.64}                & 84.34                                 \\ \hline
\end{tabular}
}
\caption{Summarised is the classification performance when the number of features for sequential forward selection and sequential backward elimination are fixed. Bold depict largest values for the column.}
\label{tab:sfs_sbe}
\end{table*}

\begin{figure*}
    \centering
    \includegraphics[scale=0.95]{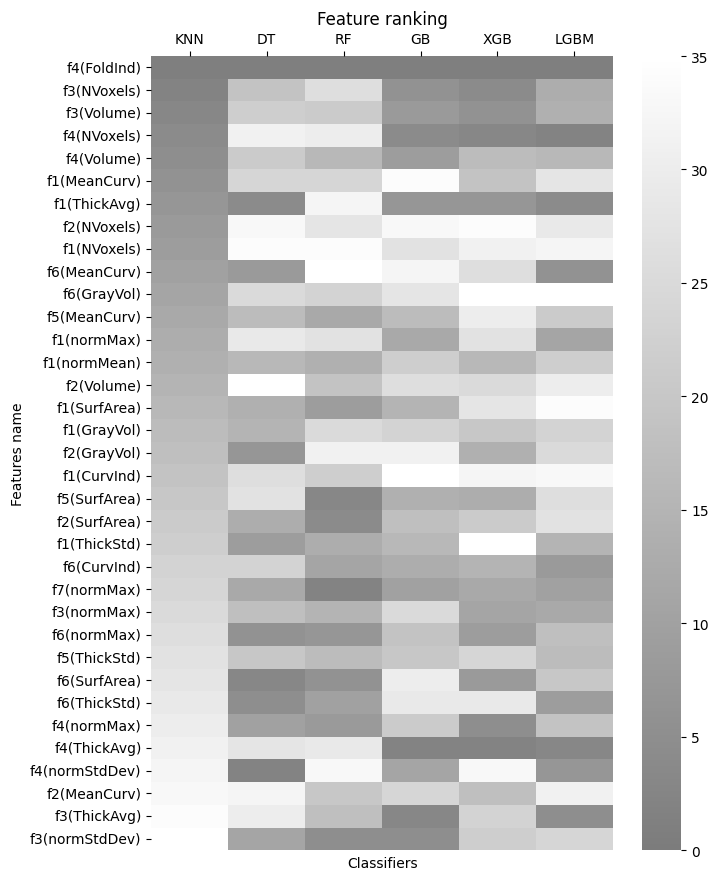}
    \caption{Feature ranking based on classification accuracy. The lower number and dark colour represent higher rank whereas higher number and brighter colour represent low rank, respectively.}
    \label{fig:feature_ranking}
\end{figure*}

\begin{table*}[]
\centering
\scalebox{0.95}{
\begin{tabular}{|c|c|l|}
\hline
\textbf{Rank} & \textbf{Feature} & \multicolumn{1}{c|}{\textbf{Interpretation}}                                                                                         \\ \hline
1             & f4(FoldInd)      & Asymmetry in cortical surface folding between the hemispheres.                                                                       \\ \hline
2             & f3(NVoxels)      & Asymmetry in the number of sub-cortical voxels between the hemispheres,   based on the number of regions above a specific threshold. \\ \hline
3             & f3(Volume)       & Asymmetry in sub-cortical volume between the hemispheres, based on the   number of regions above a specific threshold.               \\ \hline
4             & f4(NVoxels)      & Asymmetry in the number of sub-cortical voxels between the hemispheres,   based on the number of regions below a specific threshold. \\ \hline
5             & f4(Volume)       & Asymmetry in sub-cortical volume between the hemispheres, based on the   number of regions below a specific threshold.               \\ \hline
\end{tabular}
}
\caption{Selected features for the classification algorithms using KNN-based sequential forward selection approach.}
\label{tab:mri_features}
\end{table*}

\subsection{Ranking of features}
We compared sequential forward feature selection with KNN, DT, RF, GB, XGB and LGBM. Figure \ref{fig:feature_ranking} depicts the ranking of features selected using each of the algorithms and Table \ref{tab:mri_features} outlines the interpretation of the top-ranked features. To aid visualisation of the result, Figure  \ref{fig:feature_ranking} depicts feature rankings compared with the KNN rank from highest (i.e., 1) to lowest (i.e., 35). In Figure  \ref{fig:feature_ranking}, the rows correspond to NeuroMorphix ($f_1$ to $f_7$) features and FreeSurfer parameters with the row order reflecting the feature rank using KNN for selection. The gray scale reflects the feature rank obtained with each selection algorithm from 1 (darkest) to 35 (lightest).

Feature $f_4(FoldInd)$ was the highest-ranked feature for all feature selection algorithms tested. Asymmetry in cortical region size (measured by $NVoxels$ and $Volume$) also ranked highly, along with cortical thickness ($ThickAvg$) across selection methods. These findings are in keeping with  pathological changes in the cerebral cortex associated with epilepsy. Interhemispheric differences in regional $\text{T}_1$-weighted MRI intensity did not appear to be relevant for distinguishing between seizure recurrence and non-recurrence groups.

\subsection{Cross-validation}
Cross-validation was performed to validate the machine learning models, allowing us to assess the stability of prediction accuracy, generalisability to an unknown dataset, and overfitting. In Table \ref{tab:cross-val}, 5-fold cross-validation results are provided for models using sequential forward selection and feature subsets comprising 5, 7, 15, and 30 features. The optimal feature subset contained 5 features. For larger subsets, additional features were randomly selected from the remaining features.

\begin{table*}[]
\centering
\scalebox{1}{
\begin{tabular}{|c|llllllllllll|}
\hline
\multirow{3}{*}{\textbf{Number of features}} & \multicolumn{12}{c|}{\textbf{5-fold cross-validation accuracy}}                                                                                                                                                                                                                                                                                                                                                                                                       \\ \cline{2-13} 
                                           & \multicolumn{2}{c|}{\textbf{KNN}}                                         & \multicolumn{2}{c|}{\textbf{DT}}                                          & \multicolumn{2}{c|}{\textbf{RF}}                                          & \multicolumn{2}{c|}{\textbf{GB}}                                          & \multicolumn{2}{c|}{\textbf{XGB}}                                         & \multicolumn{2}{c|}{\textbf{LGBM}}                                        \\ \cline{2-13} 
                                           & \multicolumn{1}{c|}{\textit{Mean}} & \multicolumn{1}{c|}{\textit{StdDev}} & \multicolumn{1}{c|}{\textit{Mean}} & \multicolumn{1}{c|}{\textit{StdDev}} & \multicolumn{1}{c|}{\textit{Mean}} & \multicolumn{1}{c|}{\textit{StdDev}} & \multicolumn{1}{c|}{\textit{Mean}} & \multicolumn{1}{c|}{\textit{StdDev}} & \multicolumn{1}{c|}{\textit{Mean}} & \multicolumn{1}{c|}{\textit{StdDev}} & \multicolumn{1}{c|}{\textit{Mean}} & \multicolumn{1}{c|}{\textit{StdDev}} \\ \hline
5                                          & \multicolumn{1}{l|}{81.38}         & \multicolumn{1}{l|}{10.76}           & \multicolumn{1}{l|}{84.83}         & \multicolumn{1}{l|}{11.46}           & \multicolumn{1}{l|}{70.69}         & \multicolumn{1}{l|}{11.07}           & \multicolumn{1}{l|}{84.14}         & \multicolumn{1}{l|}{12.87}           & \multicolumn{1}{l|}{85.17}         & \multicolumn{1}{l|}{11.67}           & \multicolumn{1}{l|}{83.10}         & 12.40                                \\ \hline
7                                          & \multicolumn{1}{l|}{81.38}         & \multicolumn{1}{l|}{5.37}            & \multicolumn{1}{l|}{78.28}         & \multicolumn{1}{l|}{6.41}            & \multicolumn{1}{l|}{75.17}         & \multicolumn{1}{l|}{3.55}            & \multicolumn{1}{l|}{80.00}         & \multicolumn{1}{l|}{9.79}            & \multicolumn{1}{l|}{83.10}         & \multicolumn{1}{l|}{7.51}            & \multicolumn{1}{l|}{83.79}         & 7.44                                 \\ \hline
15                                         & \multicolumn{1}{l|}{84.48}         & \multicolumn{1}{l|}{3.08}            & \multicolumn{1}{l|}{75.17}         & \multicolumn{1}{l|}{4.44}            & \multicolumn{1}{l|}{78.28}         & \multicolumn{1}{l|}{3.20}            & \multicolumn{1}{l|}{79.31}         & \multicolumn{1}{l|}{4.50}            & \multicolumn{1}{l|}{85.86}         & \multicolumn{1}{l|}{4.28}            & \multicolumn{1}{l|}{84.83}         & 2.75                                 \\ \hline
30                                         & \multicolumn{1}{l|}{85.17}         & \multicolumn{1}{l|}{3.20}            & \multicolumn{1}{l|}{82.07}         & \multicolumn{1}{l|}{4.70}            & \multicolumn{1}{l|}{85.86}         & \multicolumn{1}{l|}{4.80}            & \multicolumn{1}{l|}{83.45}         & \multicolumn{1}{l|}{4.44}            & \multicolumn{1}{l|}{89.31}         & \multicolumn{1}{l|}{3.68}            & \multicolumn{1}{l|}{90.69}         & 4.44                                 \\ \hline
\end{tabular}
}
\caption{The 5-fold cross-validation results for accuracy using different number of sequentially forward selected features for each classification method.}
\label{tab:cross-val}
\end{table*}

All models were fairly robust. While classification accuracy did not necessarily improve with larger feature subsets, the standard deviation of accuracy fell, suggesting increased robustness. Trade-offs between accuracy and computational efficiency are considerations for robust classification model design.

\subsection{Generalisability of classification models}
To assess generalisability, we considered using only the top five features from the sequential forward selection algorithm. Table \ref{tab:mri_results} reports the training and testing performance of each classification model. For training, classification performance was excellent for all models, with the AUROC ranging from 85.7\% to 90.3\%. The best training performance for AUROC, accuracy, sensitivity, specificity, and $\text{F}_1$-score were 89.9\% (XGB), 84.2\% (DT), 100\% (DT, GB, LGBM), 69.3\% (KNN, RF), and 86.4\% (DT). The best performance on the test dataset for AUROC, accuracy, sensitivity, specificity, and $\text{F}_1$-score were 92.6\% (GB), 88.5\% (DT, RF, GB), 100\% (DT, GB, LGBM), 79.6\% (RF, XGB) and 89.6\% (DT, GB). This high level of consistency in performance between the training and testing datasets confirms that the classification models generalise well for this problem.

\begin{table*}[]
\centering
\scalebox{1}{
\begin{tabular}{|l|ll|ll|ll|ll|ll|ll|}
\hline
\multicolumn{1}{|c|}{\multirow{2}{*}{\textbf{Metrics}}} & \multicolumn{2}{c|}{\textbf{KNN}}                                        & \multicolumn{2}{c|}{\textbf{DT}}                                         & \multicolumn{2}{c|}{\textbf{RF}}                                         & \multicolumn{2}{c|}{\textbf{GB}}                                         & \multicolumn{2}{c|}{\textbf{XGB }}                                        & \multicolumn{2}{c|}{\textbf{LGBM }}                                       \\ \cline{2-13} 
\multicolumn{1}{|c|}{}                                  & \multicolumn{1}{c|}{\textbf{Train}} & \multicolumn{1}{c|}{\textbf{Test}} & \multicolumn{1}{c|}{\textbf{Train}} & \multicolumn{1}{c|}{\textbf{Test}} & \multicolumn{1}{c|}{\textbf{Train}} & \multicolumn{1}{c|}{\textbf{Test}} & \multicolumn{1}{c|}{\textbf{Train}} & \multicolumn{1}{c|}{\textbf{Test}} & \multicolumn{1}{c|}{\textbf{Train}} & \multicolumn{1}{c|}{\textbf{Test}} & \multicolumn{1}{c|}{\textbf{Train}} & \multicolumn{1}{c|}{\textbf{Test}} \\ \hline
\textbf{AUROC}                                          & \multicolumn{1}{l|}{86.57}          & 89.01                              & \multicolumn{1}{l|}{92.24}          & 88.74                              & \multicolumn{1}{l|}{86.82}          & 88.66                              & \multicolumn{1}{l|}{90.22}          & 92.55                              & \multicolumn{1}{l|}{92.16}          & 89.90                              & \multicolumn{1}{l|}{89.50}          & 87.55                              \\ \hline
\textbf{Accuracy}                                       & \multicolumn{1}{l|}{85.17}          & 82.76                              & \multicolumn{1}{l|}{86.90}          & 88.51                              & \multicolumn{1}{l|}{82.41}          & 88.51                              & \multicolumn{1}{l|}{85.52}          & 88.51                              & \multicolumn{1}{l|}{86.90}          & 87.36                              & \multicolumn{1}{l|}{83.79}          & 86.21                              \\ \hline
\textbf{Sensitivity}                                    & \multicolumn{1}{l|}{99.31}          & 88.37                              & \multicolumn{1}{l|}{100.00}         & 100.00                             & \multicolumn{1}{l|}{91.03}          & 90.70                              & \multicolumn{1}{l|}{100.00}         & 100.00                             & \multicolumn{1}{l|}{100.00}         & 95.35                              & \multicolumn{1}{l|}{100.00}         & 100.00                             \\ \hline
\textbf{Specificity}                                    & \multicolumn{1}{l|}{71.03}          & 77.27                              & \multicolumn{1}{l|}{73.79}          & 77.27                              & \multicolumn{1}{l|}{73.79}          & 79.55                              & \multicolumn{1}{l|}{71.03}          & 77.27                              & \multicolumn{1}{l|}{73.79}          & 79.55                              & \multicolumn{1}{l|}{67.59}          & 72.73                              \\ \hline
\textbf{F1}                                            & \multicolumn{1}{l|}{87.01}          & 83.53                              & \multicolumn{1}{l|}{88.41}          & 89.58                              & \multicolumn{1}{l|}{83.81}          & 85.71                              & \multicolumn{1}{l|}{87.30}          & 89.58                              & \multicolumn{1}{l|}{88.41}          & 88.17                              & \multicolumn{1}{l|}{86.05}          & 87.76                              \\ \hline
\end{tabular}
}
\caption{Classification performance of each algorithm for seizure recurrence prediction tabulated for the training and testing datasets.}
\label{tab:mri_results}
\end{table*}

\section{Discussion}
NeuroMorphix produces whole-brain asymmetry features based on  seven distinct mappings involving MRI-based morphological parameters of brain regions. In essence, NeuroMorphix converts image-derived parameters to whole-brain asymmetry features. We found that a subset of NeuroMorphix features were highly predictive of seizure recurrence after a first seizure. 

\subsection{Relevance of features}
An important aspect of the NeuroMorphix features is that they allow the types of brain asymmetries relevant to the cohort studied to be identified. We did this by ranking feature importance for seizure recurrence prediction, as in Figure \ref{fig:feature_ranking}. Cortical features ranked more highly than subcortical regions. This has face validity because changes in cortical thickness (see $f_1(\text{ThickAvg})$ in Figure 3), folding (i.e., $f_4(\text{FoldInd})$), and gray-white matter junction blurring ($\text{GrayVol}$ for cortical; $\text{NVoxels}$ and $\text{Volume}$ for subcortical) are observed in some epileptogenic pathologies such as focal cortical dysplasia \cite{kabat2012focal, widdess2006neuroimaging, chen2018gray, taylor1971focal}. While, in some instances, these pathologies may be detected by visual inspection of clinical brain MRI scans \cite{ruber2018mri}, they may also be MRI-negative \cite{sinha2021mapping}. The importance and power of NeuroMorphix features may lie in their sensitivity to subtle MRI changes easily missed on visual inspection of the images by a radiologist. Prediction of seizure recurrence using NeuroMorphix features achieved a high accuracy, raising the possibility for future clinical translation to guide the treatment of patients after their first seizure to reduce the risk from further seizures without unnecessary drug treatment. A better understanding of the anatomical correlates of the changes in NeuroMorphix features associated with seizure recurrence may also provide new insights into underlying mechanisms.

\subsection{NeuroMorphix features}
Each NeuroMorphix feature is a holistic representation of a specific type of morphological brain asymmetry based on region-specific parameters. For other applications, other features could be added to the current set; this would necessitate further evaluation in the relevant clinical cohorts. We did not optimise thresholds required by some features. In particular, the outlier count features, $f_3$ and $f_4$, require $\varepsilon$ to be specified.  We chose one standard deviation from the mean as $\varepsilon$, but it is possible that a different choice may alter the ranks of $f_3$ and $f_4$. It is thus plausible to suggest the choice of threshold used for $f_3$ and $f_4$ in NeuroMorphix may be application specific. 

\subsection{Robustness of FreeSurfer parameters}
We used FreeSurfer to segment and produce region-specific parameters. Although NeuroMorphix features do not rely on the use of FreeSurfer, they do require a method of segmenting and parameterising brain regions. We chose FreeSurfer because segmentation and parameterisation are integrated into the package, and it is widely used and shown to be robust.

$\text{T}_1$-weighted clinical MRI scans were used as the input to FreeSurfer. The use of multiple MRI inputs (such as a combination of $\text{T}_1$- and $\text{T}_2$-weighted images) improves FreeSurfer segmentation performance \cite{heinen2016robustness} and improves cortical measurements, such as thickness \cite{lusebrink2013cortical}. We used images resampled to $1mm^3$ isotropic resolution as inputs, a routine choice with FreeSurfer; data of different resolution can pose challenges. The use of additional MRI contrasts, the influence of the resolution of acquired images, and the acquisition of data on different scanners were beyond the scope of this study but are potential avenues for future investigation.

\subsection{Other considerations}
Class imbalance is an important aspect of classification studies. In addition to SMOTE, we examined other approaches to upsampling the minority class, such as random oversampling and found SMOTE to perform the best (data not presented here). This finding is in accordance with previous work demonstrating that SMOTE is a robust approach to the class imbalance problem \cite{chawla2002smote}.

The amount of data available and the number of NeuroMorpix features created are also important considerations. For this reason, we evaluated both sequential forward selection and sequential backward elimination. The former allows smaller NeuroMorphix feature sets to be considered (refer to Figure \ref{fig:sfs}), which is advantageous when data are limited. Additionally, NeuroMorphix features may produce a level of redundancy,  and ranking must be interpreted carefully. For example, the FreeSurfer parameter $\text{FoldInd}$ was only relevant to $f_4$, but $\text{NVoxels}$ ranked highly with $f_1$ to $f_4$ (refer to Figure \ref{fig:feature_ranking}). This suggests that region-specific volume asymmetries are highly relevant to seizure recurrence and measurable using a range of metrics.

Subsequent studies may find it useful to manually reduce the set of NeuroMorphix features instead of relying solely on an automated approach for feature selection. This should, however, be tested experimentally in the context of specific clinical cohorts. Such an approach may be of value when there is prior knowledge about the importance of specific region-specific asymmetry.

\subsection{NeuroMorphix beyond epilepsy}
In some neurodegenerative conditions such as Alzheimer’s disease, frontotemporal dementia, and Parkinson’s disease, neuropathological changes do not affect the cerebral hemispheres symmetrically. Gray matter loss in Alzheimer’s disease occurs more rapidly in the left hemisphere than in the right \cite{thompson2003dynamics}. In early Parkinson’s disease, motor symptoms are characteristically asymmetrical, with the left nigrostriatal system being more vulnerable early in the disease. Claassen et al. \cite{claassen2016cortical} observed that cortical atrophy in Parkinson’s disease tended to occur first in the left hemisphere and in frontotemporal dementia, asymmetry in disease burden between the cerebral hemispheres is thought to contribute to the clinical heterogeneity of the disorder \cite{irwin2018asymmetry}. Asymmetric atrophy of the gray matter also occurs in amyotrophic lateral sclerosis \cite{devine2015exposing}, and in multiple sclerosis, MRI-based texture and diffusion tensor parameters may show interhemispheric differences \cite{savio2015hemispheric}. NeuroMorphix features may have a role in detecting and characterising these asymmetries early in the disease and tracking their evolution as a means of furthering the understanding of selective regional and hemispheric vulnerability.

\section{Conclusion}
We developed NeuroMorphix, a framework to produce a set of features sensitive to interhemispheric differences in region-specific parameters generated from clinical MRI scans of the brain. We tested the ability of state-of-the-art classification models to use NeuroMorphix features to predict seizure recurrence in patients after their first unprovoked seizure. High prediction accuracy was achieved. Our approach also lends itself to a feature ranking nomenclature and a framework for interpreting the morphological asymmetries associated with seizure recurrence. The methods proposed here may also be used to detect, characterise, and track hemispheric asymmetries in a range of other brain disorders.

\section*{Acknowledgment}
This work was supported by Australian Research Council funded Training Centre for Innovation in Biomedical Imaging Technology (CIBIT; IC170100035) and Royal Brisbane and Women’s Hospital (Brisbane, Australia) Foundation grant, both awarded in 2018.

\bibliographystyle{ieeetr}
\bibliography{Bibliography.bib}
\end{document}